# *S*-shaped suppression of the superconducting transition temperature in Cu-intercalated $NbSe_2$

Huixia Luo[1,2*], Judyta Strychalska–Nowak[3], Jun Li[4], Jing Tao[4], Tomasz Klimczuk[3], and Robert J. Cava[1*]

[1] *Department of Chemistry, Princeton University, Princeton, New Jersey 08544, USA.*

*Email: luohx7@mail.sysu.edu.cn or rcava@princeton.edu*

[2] *School of Materials Science and Engineering, Sun Yat-Sen University, No. 135, Xingang Xi Road, Guangzhou, 510275, P. R. China*

[3] *Faculty of Applied Physics and Mathematics, Gdansk University of Technology, Narutowicza 11/12, 80–233 Gdansk, Poland*

[4] *Condensed Matter Physics and Materials Science Departments, Brookhaven National Laboratory, Upton, New York 11973, USA*

## ABSTRACT

2H-$NbSe_2$ is the prototype and most frequently studied of the well-known transition metal dichalcogenide (TMDC) superconductors. Widely acknowledged to be a conventional superconductor, its transition temperature to the superconducting state ($T_c$) is 7.3 K – a $T_c$ that is substantially higher than those seen for the majority of TMDCs, where $T_c$s between 2 and 4 K are the norm. Here we report the intercalation of Cu into 2H-$NbSe_2$ to make $Cu_xNbSe_2$. As is typically found when chemically altering an optimal superconductor, $T_c$ decreases with increasing x, but the way that $T_c$ is suppressed in this case is unusual – an *S*-shaped character is observed, with an inflection point near x = 0.03 and, at higher x, a leveling off of the $T_c$ near 3 K – down to the usual value for a layered TMDC. Electronic characterization reveals corresponding *S*-like behavior for many of the materials parameters that influence $T_c$. To illustrate its character, the superconducting phase diagram for $Cu_xNbSe_2$ is contrasted to those of $Fe_xNbSe_2$ and $NbSe_{2-x}S_x$.

## INTRODUCTION

Transition metal dichalcogenides (TMDCs) have been studied for decades due to the rich electronic properties that arise due to their low structural dimensionality. These systems share the $MX_2$ formula, where M is a transition metal (M = Ti, Zr, Hf, V, Nb, Ta, Mo, W or Re), and X is a chalcogen (X = S, Se, or Te); [1-4] the structures are made from stacking X-M-X layers in repeating patterns, with Van der Waals (VdW) bonding between the layers. 2H-$NbSe_2$ was one of the earliest layered TMDC materials known to superconduct, with a critical temperature $T_c$ ~7.3 K. This $T_c$ is significantly higher than is encountered for the many other TMDC superconductors known, where $T_c$ is commonly in the 2 - 4 K range. 2H-$NbSe_2$ also hosts a quasi–two-dimensional incommensurate charge density wave (ICDW) with $T_{CDW}$ ~ 33 K. [5] It is the most studied of the layered transition metal dichalcogenide superconductors, with almost countless experimental and theoretical papers focusing on its behavior over the past 50 years. (see, e.g.[6-14]) The relation between Fermi surface nesting and its superconductivity is still under debate, [12] for example, even though 2H-$NbSe_2$ has been considered a conventional superconductor for decades. [13] In addition, due to recent new concepts, such as d-wave pairing in cuprates [14] and two-gap superconductivity in $MgB_2$, [15-17] studies of the superconducting order parameters of 2H-$NbSe_2$ have recently increased in number. [18-21] Many recent experiments performed on 2H-$NbSe_2$, including specific heat, [18] thermal conductivity, [19] magnetization, [20] penetration depth, [21] tunneling spectroscopy [22-24] and angle-resolved photoemission spectroscopy (ARPES) [25-29] agree that more than one energy scale is important for the superconductivity. Evidence for strong gap anisotropy in 2H-$NbSe_2$ thus appears to be well established, but it remains under discussion whether this is a result of there being different superconducting gaps on different Fermi surface sheets, or whether it originates elsewhere.

Here we report the results of the intercalation of Cu into 2H-$NbSe_2$ to form $Cu_xNbSe_2$ [30] in the doping range of $0 \leq x \leq 0.09$, where the 2H structure (see **Fig. 1A**) is maintained; above x = 0.09 the material is multiple phase and deductions about $T_c$ cannot reliably be made. The $T_c$ decreases with increasing Cu content in $Cu_xNbSe_2$, as is commonly found when introducing "impurities" into optimal superconductors, but the way that superconductivity is suppressed is unusual. An *S*-shaped character is observed in $T_c$ vs. x, with an inflection point near x = 0.03 and a leveling off, at higher x, of the abnormally high 7 K $T_c$ in the pure phase to a $T_c$ value near 3 K – back down to where it is commonly observed for a layered TMDC. The materials are characterized through measurements of their resistivities, critical fields, magnetic susceptibilities and heat capacities, which reveal a corresponding S-shaped behavior in the electronic properties of the system, most notably the electronic contribution to the specific heat and the electron-phonon coupling parameter. CDWs compete with superconductivity for stability at low temperatures in the layered TMDCs and thus the effect of Cu-intercalation on the CDW is of interest to obtain a fuller picture of the electronic system. Our temperature-dependent electron diffraction study reveals a minor change in the *q*-vector of the CDW on Cu intercalation, degradation but not destruction of the coherence of the CDW, especially in the direction perpendicular to the layers, and also the extension of CDW fluctuations, evidenced by diffuse scattering, to room temperature. Comparison of the superconducting phase diagram for $Cu_xNbSe_2$ to those for the other doped 2H-$NbSe_2$ materials $Fe_xNbSe_2$ and $NbSe_{2-x}S_x$ illustrates its unusual character.

## EXPERIMENTAL SECTION

Polycrystalline samples of $Cu_xNbSe_2$ were synthesized in two steps by solid state reaction. First, the mixtures of high-purity, cleaned fine powders of Cu (99.9%), Nb (99.9%), and Se (99.999%) in the appropriate stoichiometric ratios were heated in sealed evacuated silica glass tubes at a rate of 1 $^{o}$C/min to 700 $^{o}$C and held there for 120 h. Subsequently, the as-prepared powders were reground, re-pelletized, and sintered again, by heating at a rate of 3 $^{o}$C/min to 700 $^{o}$C and holding there for 48 h. The $NbSe_{2-x}S_x$ samples were prepared in the same way. The identity and phase purity of the samples were determined by powder X-ray diffraction (PXRD) using a Bruker D8 Advance ECO with Cu Kα radiation and a LYNXEYE-XE detector. To determine the unit cell parameters, profile fits were performed on the powder diffraction data through the use of the FULLPROF diffraction suite using Thompson-Cox-Hastings pseudo-Voigt peak shapes. [31]

Measurements of the temperature dependent electrical resistivity (4 contact method), specific heat, and magnetic susceptibility of the materials were performed in a DynaCool Quantum Design Physical Property Measurement System (PPMS). There was no indication of air-sensitivity of the materials during the study. $T_c$s determined from susceptibility data were estimated conservatively: $T_c$ was taken as the intersection of the extrapolations of the steepest slope of the susceptibility in the superconducting transition region and the normal state susceptibility; for resistivities, the midpoint of the resistivity ρ(T) transitions was taken, and, for the specific heat data, the critical temperatures obtained from the equal area construction method were employed.

The temperature dependent electron diffraction experiments were carried out on a JEOL 2100F transmission electron microscope equipped with a Gatan liquid helium sample stage. Coherent electron diffraction patterns were recorded by a CCD camera and obtained from the same area in a single-crystal domain for each sample throughout the thermal process. For each sample, electron diffraction patterns acquired from several single-crystal domains were all consistent, with typical results shown here.

## Results and Discussion

**Figs. 1 B** and **C** show the powder X-ray diffraction patterns and unit cell parameters for $Cu_xNbSe_2$ ($0 \leq x \leq 0.09$). The results show that a single phase solid solution is indeed formed. The solubility limit for intercalated Cu in 2H-$NbSe_2$ is x = 0.09. At higher Cu

contents, the cubic $Cu_3NbSe_4$ phase is found as an impurity. Within the solid solution, the unit cell parameters *a* and *c* both increase linearly with increasing Cu content, in a Vegard's law type behavior: *a* increases linearly from 3.4432(4) Å (x = 0) to 3.4507(1) Å (x = 0.09) and *c* increases linearly from 12.5409(5) (x = 0) to 12.6277(4) (x = 0.09). The increase of *c* with increasing Cu content is a signature that is characteristic of 3d metal intercalation in TMDCs.[32] The detailed changes of the crystallographic cell in the $NbSe_{2-x}S_x$ system are not the subject of this study and are not presented.

We next consider the temperature dependence of the normalized ($\rho/\rho_{300K}$) resistivities of both $Cu_xNbSe_2$ and $NbSe_{2-x}S_x$. Although careful interpretation of resistivities necessitates the use of data obtained on single crystals, consideration of the data on polycrystalline samples can provide some basic insights. **Figs. 2A** and 2B thus show the temperature dependence of the normalized electrical resistivities ($\rho/\rho_{300K}$) for polycrystalline samples of both $NbSe_{2-x}S_x$ and $Cu_xNbSe_2$. The samples in both cases show a metallic temperature dependence ($d\rho/dT > 0$) in the temperature region of 8 – 300 K. However, the relative resistances of the $Cu_xNbSe_2$ samples decrease substantially less with temperature than those of $NbSe_{2-x}S_x$; the residual resistivity ratios ((resistivity at 300 K)/(resistivity just above $T_c$)) for the polycrystalline samples are, for example, ~ 29 for $NbSe_2$, ~ 9 for $NbSe_{1.9}S_{0.1}$, and ~ 2 for $Cu_{0.09}NbSe_2$. The factor of 15 differences between $NbSe_2$ and $Cu_{0.09}NbSe_2$ suggests that Cu may be an electronically disruptive dopant in $NbSe_2$. At low temperatures (see **Fig. 2C** and **2D**), a clear, sharp drop of $\rho(T)$ is observed in all the $Cu_xNbSe_2$ and $NbSe_{2-x}S_x$ samples, signifying the onset of superconductivity at low temperatures. $T_c$ decreases with higher doping content in both cases. This trend is also clearly seen in the susceptibility data (**Fig. 2E** and **2F**) - the onset of the negative magnetic susceptibility signaling the superconducting state shifts systematically to lower temperatures with increasing x for both $Cu_xNbSe_2$ and $NbSe_{2-x}S_x$.

More detailed information on the electronic properties and superconductivity of the $Cu_xNbSe_2$ solid solution was obtained from specific heat measurements. **Fig. 3A** shows the temperature dependence of the zero-field specific heat, $C_p/T$ versus T, for selected $Cu_xNbSe_2$ samples. The figure shows that all the materials display a large specific heat jump at $T_c$, an indication of bulk superconductivity. The superconducting transition temperatures are in excellent agreement with the $T_c$s determined from the $\rho(T)$ and $\chi(T)$ measurements. The normal state specific heats at low temperatures in the presence of a magnetic field large enough to suppress the superconductivity obey the relation of $C_p = \gamma T + \beta T^3$, where $\gamma$ and $\beta$ describe the electronic and phonon contributions to the heat capacity, respectively. By fitting the data obtained in the 4T and 8T applied field (**Fig. 3B**), we obtain the electronic specific heat coefficients ($\gamma$) and phonon specific heat coefficients $\beta$. The normalized specific heat jump values $\Delta C/\gamma T_c$ thus obtained from the data in **Figs. 3A** and **3B** range from 2.04 for 2H-$NbSe_2$, to 1.68 for $Cu_{0.07}NbSe_2$, respectively. These are all higher than the Bardeen-Cooper-Schrieffer (BCS) weak-coupling limit value (1.43), and clearly decrease with increasing x. Using the fitted values of β, we estimate the Debye temperatures by the relation $\theta_D = (12\pi 4nR/5\beta)^{1/3}$, where n is the number of atoms per formula unit, and R is the gas constant. The results (**Fig. 3C**) show that the Debye temperatures increase modestly with increasing Cu content in $Cu_xNbSe_2$ as the lattice becomes stiffer when some of the Se-Se VdW bonds are replaced by Se-Cu-Se bonds. These data are summarized in **Table 1**.

The dependence on x of the superconducting transition temperature and important electronic parameters for $Cu_xNbSe_2$ determined from the specific heat data are summarized in **Fig. 4**. Using the Debye temperature ($\theta_D$), the critical temperature $T_c$, and assuming that the electron-phonon coupling constant ($\lambda_{ep}$) can be calculated from the inverted McMillan formula:[33]

$$\lambda_{ep} = \frac{1.04 + \mu^* \ln\left(\frac{\theta_D}{1.45T_C}\right)}{(1 - 0.62\mu^*)\ln\left(\frac{\theta_D}{1.45T_C}\right) - 1.04}$$

the values of $\lambda_{ep}$ obtained range from 0.81 for 2H-$NbSe_2$ to 0.55 for $Cu_{0.09}NbSe_2$ (**Table 1**). These values suggest strong coupling superconductivity. With the Sommerfeld parameter ($\gamma$) and the electron-phonon coupling ($\lambda_{ep}$), the electron density of states at the Fermi level ($N(E_F)$) can be calculated from $N(E_F) = \frac{3}{\pi^2 k_B^2 (1+\lambda_{ep})}\gamma$. This yields values that range from $N(E_F)$ = 4.08 states/eV f.u. for $NbSe_2$ to $N(E_F)$ = 2.39 states/eV f.u. for $Cu_{0.09}NbSe_2$ (**Table 1**). The density of electronic states at the Fermi energy therefore clearly decreases when more Cu intercalates into 2H-$NbSe_2$. It can be seen in **Fig. 4** that the electronic parameters derived from the specific heat for $Cu_xNbSe_2$ show an *S*-shaped character that corresponds to that for the superconducting transition temperatures.

The superconducting transitions for selected $Cu_xNbSe_2$ samples were further examined through temperature dependent measurements of the electrical resistivity and magnetization under applied magnetic field, with the goal of determining the critical fields at 0 K, $\mu_0H_{c1}(0)$, $\mu_0H_{c2}(0)$ and $\mu_0H_c(0)$. First we consider the resistivity measurements employed to determine $\mu_0H_{c2}(0)$. The ρ(T,H) data obtained for $Cu_xNbSe_2$ (x = 0, 0.05) are shown as an example in **Figs. 5A** and **B**. Based on the $T_c$ determined resistively under different magnetic fields, the upper critical field values, $\mu_0H_{c2}$, are plotted vs. temperature in **Fig. 5C**. A clear linear dependence of $\mu_0H_{c2}$ vs. T is seen near $T_c$ for all samples: the solid lines through the data show the best linear fits. The initial slopes ($dH_{c2}/dT$) for $Cu_xNbSe_2$ are shown in **Table 1**. From this data we estimate the zero temperature upper critical fields (upper inset **Fig. 5F**) to range from 9.96 T for $NbSe_2$, to $\mu_0H_{c2}$ = 3.72T for $Cu_{0.07}NbSe_2$, using the Werthamer-Helfand-Hohenberg (WHH) expression for the dirty limit superconductivity, $\mu_0H_{c2} = -0.693T_c\,(dH_{c2}/dT_c)$.[33-37] The results are summarized in **Table 1**; $\mu_0H_{c2}$ for x = 0.01 is larger than that for x = 0, likely due to vortex pinning - it then decreases with increasing x. The Pauli limiting field for $Cu_xNbSe_2$ was estimated from $\mu_0H^P = 1.86T_c$. The obtained values of $\mu_0H^P$ are only slightly larger than estimated $\mu_0Hc_2$. Finally, using $\mu_0 H_{c2} = \frac{\phi_0}{2\pi\xi_{GL}^2}$, where $\phi_o$ is the quantum of flux, the Ginzburg-Laudau coherence length ($\xi_{GL}(0)$) can be estimated to range from ~ 5.3 nm for $Cu_{0.01}NbSe_2$, to ~ 10.1 nm for $Cu_{0.07}NbSe_2$ (**Table 1**).

To determine $\mu_0H_{c1}(0)$, the superconducting transition for selected $Cu_xNbSe_2$ samples was further examined through temperature dependent measurements of the magnetization under increasing applied magnetic field M(H). The main panel of **Fig. 5D** shows the data for $Cu_{0.05}NbSe_2$, and how $\mu_0H_{c1}$ was determined, as an example. First, in order to estimate the demagnetization factor (N), low-field magnetization measurements as a function of field M(H) were performed at temperatures 1.7, 2, 2.5, 3 and 3.5 K, as shown in the main panel of **Fig. 5D**. At low magnetic fields, the experimental data can be fit with the linear formula $M_{fit} = a + bH$. Assuming that the initial linear response to a magnetic field is perfectly diamagnetic ($dM/dH = -1/4\pi$) for these bulk superconductors, we obtained N, the demagnetization factor, of 0.1 ~ 0.7 (from $-4\pi\chi_V = \frac{1}{1-N}$, where $\chi_V = \frac{dM}{dH}$ is actually a fitted slope from the main panel of Fig.5 D) that is consistent with the sample shape. The M(H)-$M_{fit}$ data is plotted vs. applied magnetic field (H) in the inset of **Fig. 5D**. $\mu_0H_{c1}*$ is taken as the field where M deviates by ~ 2% above the fitted line ($M_{fit}$), as is the common practice. [38] Taking into account the demagnetization factor (N), the lower critical field at temperature T, $\mu_0H_{c1}(T)$, can then be calculated from the formula $\mu_0H_{c1}(T) = \mu_0H_{c1}*(T)/(1-N)$.[39,40] **Fig. 5E** presents $\mu_0H_{c1}$ as a function of temperature for selected $Cu_xNbSe_2$ samples. The estimation of $\mu_0H_{c1}(0)$ is then possible by fitting the $\mu_0H_{c1}(T)$ data to the formula $\mu_0H_{c1}(T) = \mu_0H_{c1}(0)[1-(T/T_c)^2]$, which is represented by the solid lines. The estimated zero-temperature lower critical fields $\mu_0H_{c1}(0)$ (see the bottom of the inset of **Fig. 5F**) range from for $Cu_xNbSe_2$ (x = 0, 0.01, 0.02, 0.03, 0.035, 0.05) are 0.0158 T for $NbSe_2$ to 0.0056 T for $Cu_{0.05}NbSe_2$. From the relation $\mu_0H_{c1}(0) = (\Phi_0/4\pi\lambda^2)\ln(\lambda_{GL}/\xi_{GL})$ we numerically find another important superconducting parameter - the magnetic penetration depth $\lambda_{GL}$. This parameter ranges from 191 nm for $NbSe_2$ to 323 nm for $Cu_{0.05}NbSe_2$. The Ginzburg-Landau parameter $\kappa_{GL} = \lambda_{GL}/\xi_{GL}$ is then calculated and confirms type-II superconductivity in $Cu_xNbSe_2$. The thermodynamic critical field $\mu_0H_c = ((\mu_0H_{c1}\ \mu_0H_{c2}/\ln\kappa)^{0.5})$ is then shown in **Fig. 5F**. $\mu_0H_c$ decreases linearly with increasing Cu content in $Cu_xNbSe_2$. These parameters are again summarized in **Table 1**.

**Fig. 6** shows the comparison of the electron diffraction patterns in the basal plane *hk0* reciprocal lattice for $NbSe_2$ and $Cu_{0.06}NbSe_2$, both at room temperature and at 10 K.

The hexagonal symmetry is clearly seen in the intense diffraction spots that arise from the basic structure for both materials at both temperatures. At 10 K, the sharp superlattice diffraction spots due to the CDW (between the intense spots from the basic lattice) are clearly seen for pure $NbSe_2$. Their measured nearly commensurate *q* vector is 0.337*a**, and they appeared sharply on cooling between 30 and 40 K, consistent with previous observations for 2H $NbSe_2$.[41,42] The measured widths of the superlattice spots suggest an in-plane coherence of 15-20 nm. Similar, but significantly different in detail, diffraction evidence for CDW formation is also clearly seen in $Cu_{0.06}NbSe_2$. In this case the spots are less sharp, an indication of a decrease in coherent diffracting volume, with a measured in-plane coherence of 10 nm or less. The q vector of the CDW has changed somewhat, to 0.370*a** for $Cu_{0.06}NbSe_2$, Finally, diffuse electron scattering characteristic of CDW formation with short coherence lengths was not clearly visible in undoped $NbSe_2$ above its 3D CDW transition ,i.e. the onset of the 3D CDW was sharp in temperature. In contrast, however, diffuse scattering persists up to room temperature in $Cu_{0.06}NbSe_2$, indicative of short coherence length CDW fluctuations in the Cu-intercalated material up to quite high temperatures. For $Cu_{0.06}NbSe_2$, no sharp transition to 3D ordering is observed, rather the $q = 0.37a^*$ diffraction spots visible at 10 K appear to grow approximately continuously out of the diffuse scattering on cooling. It is natural to ask whether the Cu intercalation disrupts the CDW coherence more in-plane than out-of-plane. This question can be addressed by looking at a larger volume of the reciprocal lattice, shown in **Fig. 7**. Far from the origin of the reciprocal lattice the curvature of the Ewald sphere allows for higher index zones (e.g. *hk1*, *hk2*, etc.) to be sampled. The results for these types of reciprocal space probes are shown for diffraction from 2H $NbSe_2$ and $Cu_{0.06}NbSe_2$ in **Fig. 7**. Again the hexagonal symmetry of the basic structure is clearly seen, as are the superlattice spots due to the CDW formation. (The bright rings of spots distant from the origin arise from the intersection of the Ewald sphere with the higher order basic structure reciprocal lattice planes.) The degradation of the intensity of the superlattice spots on going to higher zones is much larger in the case of pure $NbSe_2$ than it is for $Cu_{0.06}NbSe_2$, and therefore these diffraction patterns show that the disruption of the CDW is more pronounced perpendicular to the planes than in-plane – i.e. that the 3D CDW coherence is more disrupted than the in-plane coherence by Cu intercalation. In both cases, the CDW spots are strong in the *hk0* zone (near the origin of the reciprocal lattice), consistent with previous studies indicating that the CDW is still present in single-layer $NbSe_2$. [43]

Finally, the superconductivity phase diagram as a function of doping level for 2H-$Cu_xNbSe_2$ is summarized in **Fig. 8**. For comparison, the superconductivity phase diagram for $NbSe_{2-x}S_x$ ($0 \leq x \leq 0.1$) from the current study is included, as is the same information for $Fe_xNbSe_2$ taken from the literature.[44] The $T_c$s extracted from the three kinds of measurements performed here for $Cu_xNbSe_2$ (resistivity, magnetic susceptibility, heat capacity) are all consistent; the x dependence of $T_c$ displays an *S*-like shape in the $Cu_xNbSe_2$ system, but not in the other systems. The $T_c$ of pure 2H-$NbS_2$ is 6.5 K, [45] and therefore the very small change in $T_c$ of the sulfur doped material may not be surprising. (Work on $Cu_xNbS_2$ similar to that performed in the current study may therefore be of future interest.) In contrast, intercalated Fe may be considered as a magnetic ion, with the magnetism leading to a very rapid suppression of $T_c$ with increasing Fe content. Therefore non-magnetic chemically different Cu intercalation is naively expected to act somewhere between the Fe- and S-doped extremes, although not with an *S*-shaped behavior - there are no previously known examples of an *S*-shaped suppression of $T_c$ by substitution or doping in a single phase material.

## Conclusion

$Cu_xNbSe_2$ ($0 \le x \le 0.09$) was prepared by a solid state method. The electronic properties, including resistivity, heat capacity and critical fields were studied in detail, and indicate that copper doping suppresses the superconductivity in $NbSe_2$ in an unexpected way - the electronic properties and x dependent electronic phase diagram show that the superconducting transition temperature of Cu intercalated $NbSe_2$ shows an unusual *S*-shaped behavior. The underlying reason for this usual *S*-shape behavior of $T_c$ and the electronic characteristics that give rise to it have not been determined here. However, based on the fact that the $T_c$ is unusually high for pure $NbSe_2$ and then settles in to the usual value for TMDCs for $Cu_xNbSe_2$, it is not unreasonable to speculate that Cu doping destroys the higher energy paring channel that makes $NbSe_2$ unusual among the layered TMDCs. We speculate that this may be due to either the electron doping of the $NbSe_2$ layer that results on Cu intercalation, or to the non-magnetic disorder introduced by the Cu intercalation. The lower $T_c$ cannot, for example, simply be due to a strengthening of the competing CDW state for non-zero *x* in $Cu_xNbSe_2$, because the electron diffraction studies show that such strengthening is clearly not the case. It may, however, be that the Cu intercalation both disrupts the coherence of the CDW and simultaneously suppresses the pairing channel that gives rise to the higher $T_c$ in $NbSe_2$. Future detailed characterization experiments and theoretical treatments will be required to determine whether this speculation is indeed the case.

## Acknowledgments

The research at Princeton University on sample synthesis and structural, resistive, and susceptibility characterization was supported by the US DOE BES through grant DE-FG02-98ER45706. All work in Gdansk Poland, including specific heat measurements and their interpretation, was supported by National Science Centre (Poland), through grant UMO-2015/19/B/ST3/03127. The electron diffraction study at Brookhaven National Laboratory was supported by the DOE BES, by the Materials Sciences and Engineering Division under Contract DE-SC0012704, and through the use of the Center for Functional Nanomaterials. F. von Rohr, L. Müchler and T. Kong are thanked for useful discussions.

## Table 1 Characterization of the superconductivity in the $Cu_xNbSe_2$ family.

| *x* in $Cu_xNbSe_2$ | 0 | 0.01 | 0.02 | 0.03 | 0.035 | 0.04 | 0.05 | 0.06 | 0.07 | 0.09 |
|---|---|---|---|---|---|---|---|---|---|---|
| $T_c$ (K) | 7.16 | 7.05 | 6.04 | 5.6 | 5.2 | 4.46 | 3.8 | 3.18 | 2.9 | 2.3 |
| $\gamma$ (mJ mol$^{-1}$ K$^{-2}$) | 17.4(20) | 17.37(30) | 15.58(17) | 14.1(1) | -- | 12.11(10) | 10.5(1) | 10.4(1) | 9.7(1) | 8.8(1) |
| ß (mJ mol$^{-1}$ K$^{-4}$) | 0.56 | 0.61 | 0.56 | 0.53 | -- | 0.52 | 0.56 | 0.49 | 0.47 | 0.44 |
| $\Theta_D$ (K) | 218(16) | 212(26) | 217(12) | 222(20) | -- | 225(06) | 219(14) | 230(10) | 233(20) | 239(60) |
| $\Delta C/\gamma T_c$ | 2.04 | 2.15 | 1.94 | 1.96 | -- | 1.81 | 1.79 | 1.57 | 1.68 | 1.62 |
| $\lambda_{ep}$ | 0.81 | 0.82 | 0.76 | 0.73 | -- | 0.67 | 0.65 | 0.60 | 0.59 | 0.55 |
| $N(E_F)$ (states/eV f.u) | 4.08 | 4.17 | 3.76 | 3.45 | -- | 3.07 | 2.71 | 2.82 | 2.60 | 2.39 |
| $-dH_{c2}/dT$(T/K) | 1.95(4) | 2.39(4) | 2.29(4) | 2.15(15) | 2.01(3) | 2.25(2) | 1.45(4) | -- | 1.59(5) | -- |
| $\mu_0 H_{c2}$(T) | 9.7(2) | 11.7(2) | 9.6(2) | 8.3(6) | 7.2(1) | 7.0(1) | 3.8(1) | -- | 3.2(1) | -- |
| $\mu_0 H^P$(T) | 13.2 | 13.0 | 11.2 | 10.4 | 9.6 | 8.3 | 7.0 | | 5.4 | |
| $\xi_{GL}(0)$ (nm) | 5.8 | 5.3 | 5.9 | 6.3 | 6.7 | 6.9 | 9.3 | -- | 10.1 | -- |
| $\mu_0 H_{c1}$(T) | 0.0158 | 0.0111 | 0.0100 | 0.0080 | 0.0065 | -- | 0.0056 | -- | -- | -- |
| $\lambda_{GL}(0)$ (nm) | 191 | 237 | 248 | 280 | 314 | -- | 323 | | | |
| $\kappa_{GL}$ | 33 | 44 | 42 | 45 | 47 | -- | 35 | | | |
| $\mu_0 H_c$(mT) | 209 | 185 | 160 | 133 | 111 | -- | 78 | | | |

## Figures legends

**Fig. 1. Structural and chemical characterization of $Cu_xNbSe_2$ (A)** Flow chart for Cu intercalation of 2H-$NbSe_2$. The copper doped materials are synthesized directly from the elements (B) Powder XRD patterns (Cu Kα) for the $Cu_xNbSe_2$ samples studied ($0 \leq x \leq 0.09$). The inset shows a detail of the diffracted angle region where the effect of the increasing cell parameters with increasing Cu intercalation can clearly be seen. (C) Composition dependence of the room temperature lattice parameters for $Cu_xNbSe_2$ ($0 \leq x \leq 0.09$); standard deviations are shown when they are larger than the plotted points.

**Fig. 2. Transport characterization of the normal states and superconducting transitions for $Cu_xNbSe_2$ and $NbSe_{2-x}S_x$** (**A, B**) The temperature dependence of the resistivity ratio ($\rho/\rho_{300K}$) for polycrystalline $Cu_xNbSe_2$ ($0 \leq x \leq 0.09$) and $NbSe_{2-x}S_x$ ($0 \leq x \leq 0.1$). (**C, D**) Enlarged temperature region showing the superconducting transitions. (**E, F**) Magnetic susceptibilities for $Cu_xNbSe_2$ ($0 \leq x \leq 0.09$) and $NbSe_{2-x}S_x$ ($0 \leq x \leq 0.1$) at the superconducting transitions; applied DC fields are 20 Oe.

**Fig. 3. Heat Capacity characterization of $Cu_xNbSe_2$.** (**A**) Heat capacities through the superconducting transitions without applied magnetic field for different compositions in $Cu_xNbSe_2$ and (**B**) Heat capacities for $Cu_xNbSe_2$ for different x values at an applied magnetic field sufficiently high to fully suppress the superconductivity; data used to determine the electronic contribution to the specific heat and the Debye temperatures. (**C**) Debye temperature of $Cu_xNbSe_2$ for different x values obtained from fits to data in Figure 3B.

**Fig. 4**. **The superconducting transition temperatures for $Cu_xNbSe_2$ and the associated electronic characteristics.** (**A**) Superconducting transition temperature Tc vs x; (**B**) Electronic contribution to the specific heat, gamma, vs. x; (**C**) $\Delta C/\gamma T_c$ vs x; (D) and, finally, the electron-phonon coupling constant lambda vs x, in $Cu_xNbSe_2$. An S-like character is observed for all parameters.

**Fig. 5**. **Characterization of the critical fields of $Cu_xNbSe_2$.** (**A, B**): Low temperature resistivity at various applied fields for the examples of $NbSe_2$ and $Cu_{0.05}NbSe_2$; (**C**) The temperature dependence of the upper critical field ($\mu_0 H_{c2}$) for $Cu_xNbSe_2$; (**D**) Magnetic susceptibility at low applied magnetic field at various applied temperatures for $Cu_{0.05}NbSe_2$. The inset shows the $M$-$M_{fit}$ vs H; (**E**) The temperature dependence of the lower critical field ($\mu_0 H_{c1}$) for $Cu_xNbSe_2$ (**F**) The thermodynamic critical field vs x.

**Fig. 6. Temperature dependent electron diffraction characterization of the CDWs in $NbSe_2$ and $Cu_{0.06}NbSe_2$.** Typical electron diffraction patterns (along the [001] zone axis, i.e. in the *hk0* reciprocal lattice plane) obtained from pure $NbSe_2$ (left) and $Cu_{0.06}NbSe_2$ (right). The hexagonal symmetry is clearly seen. For both samples, patterns at room temperture (RT) and low temperatures were obtained from the same area. The vertical streaks are artifacts from the CCD camera. Kikuchi bands are seen in some of the patterns. The CDW in both cases is evidenced by lines of weaker diffraction spots between the main structure spots in the data at 10 K. For the $Cu_{0.06}NbSe_2$ material, diffuse streaks in the diffraction pattern along these lines persist to room temperature.

**Fig. 7. The 10 K electron diffraction patterns over a wider volume of reciprocal space in $NbSe_2$ and $Cu_{0.06}NbSe_2$** The intensity distributions of the CDW reflections strongly suggest the following: 1) The atomic displacements associated with the CDW modulation are longitudinal in both cases; and 2) The CDW modulation is relatively weakly correlated between layers (along the *c*-axis) for both materials. For $Cu_{0.06}NbSe_2$, the correlation of the CDW along the c-axis is weaker, based on the fact that the intensities of the CDW diffuse reflections remain strong close to the *hk1* Laue zone. Kikuchi bands and the hexagonal symmery of the patterns are clearly seen.

**Fig. 8. The superconducting phase diagram for 2H-$Cu_xNbSe_2$ compared to those of 2H $NbSe_{2-x}S_x$ from the current study and 2H-$Fe_xNbSe_2$ from Reference 44.** The usual range of Tc's for transition metal dichalcogenides is illustrated by the dashed lines. "SC" and "metallic" label the superconducting and metallic regions for $Cu_xNbSe_2$, respectively. The labels $\rho(T)$, $\chi(T)$ and $C_p(T)$ are the $T_c$s determined for the materials in the current study from resistivity, magnetic susceptibility and specific heat characterization, respectively. The limit of the $Cu_xNbSe_2$ solubility is x = 0.09; beyond that x value the materials contain multiple phases (the solubility extends to at least x = 0.1 for $NbSe_{2-x}S_x$.) Upper right inset – schematic of the crystal structure of $Cu_xNbSe_2$.

**Fig. 1.**

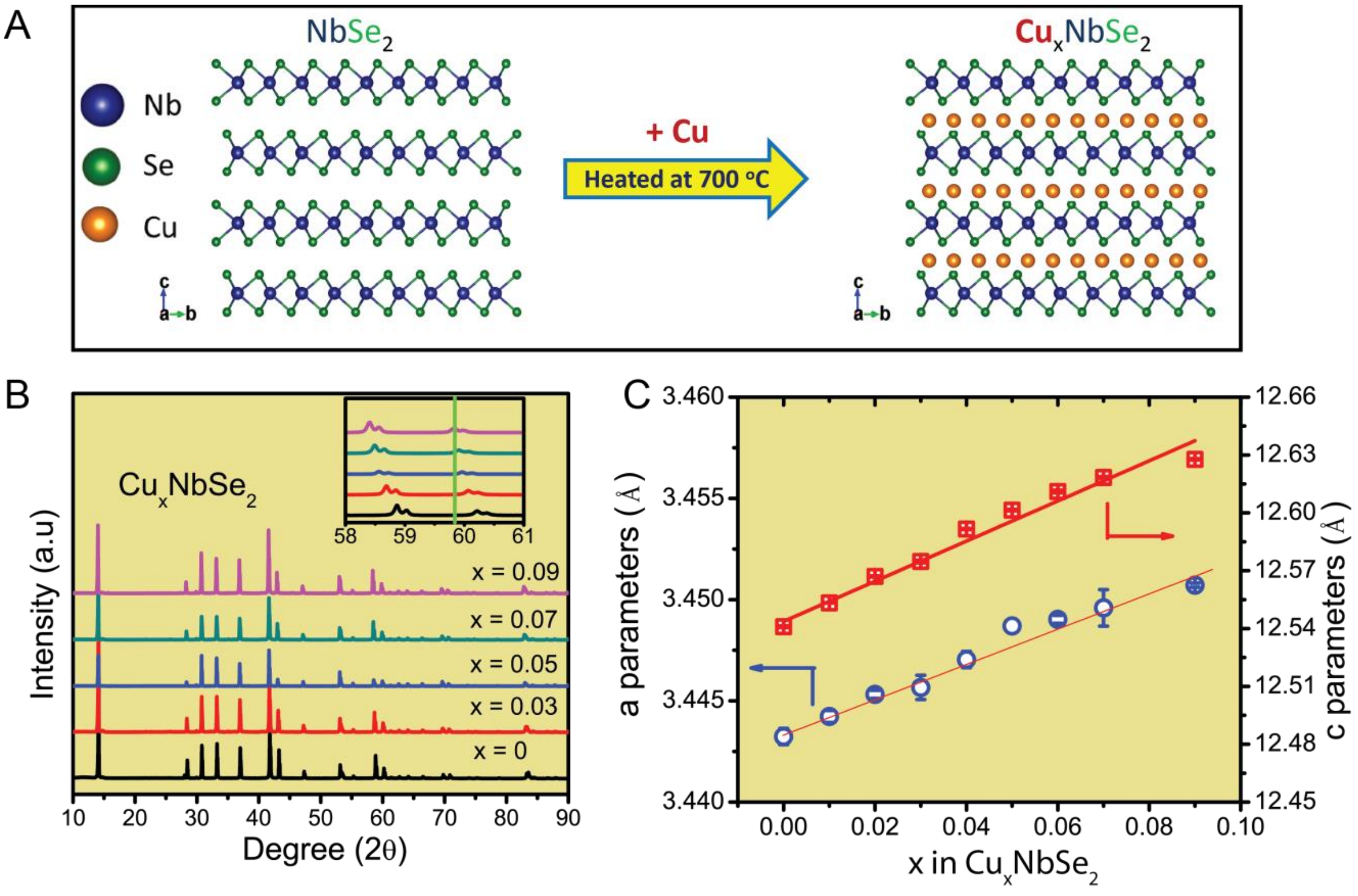
A
NbSe2
CuxNbSe2
Nb
Se
Cu
+ Cu
Heated at 700 ºC
c
a
b
B
CuxNbSe2
Intensity (a.u)
Degree (2θ)
x = 0.09
x = 0.07
x = 0.05
x = 0.03
x = 0
10 20 30 40 50 60 70 80 90
58 59 60 61
C
a parameters (Å)
c parameters (Å)
x in CuxNbSe2
3.440 3.445 3.450 3.455 3.460
12.45 12.48 12.51 12.54 12.57 12.60 12.63 12.66
0.00 0.02 0.04 0.06 0.08 0.10

**Fig. 2.**

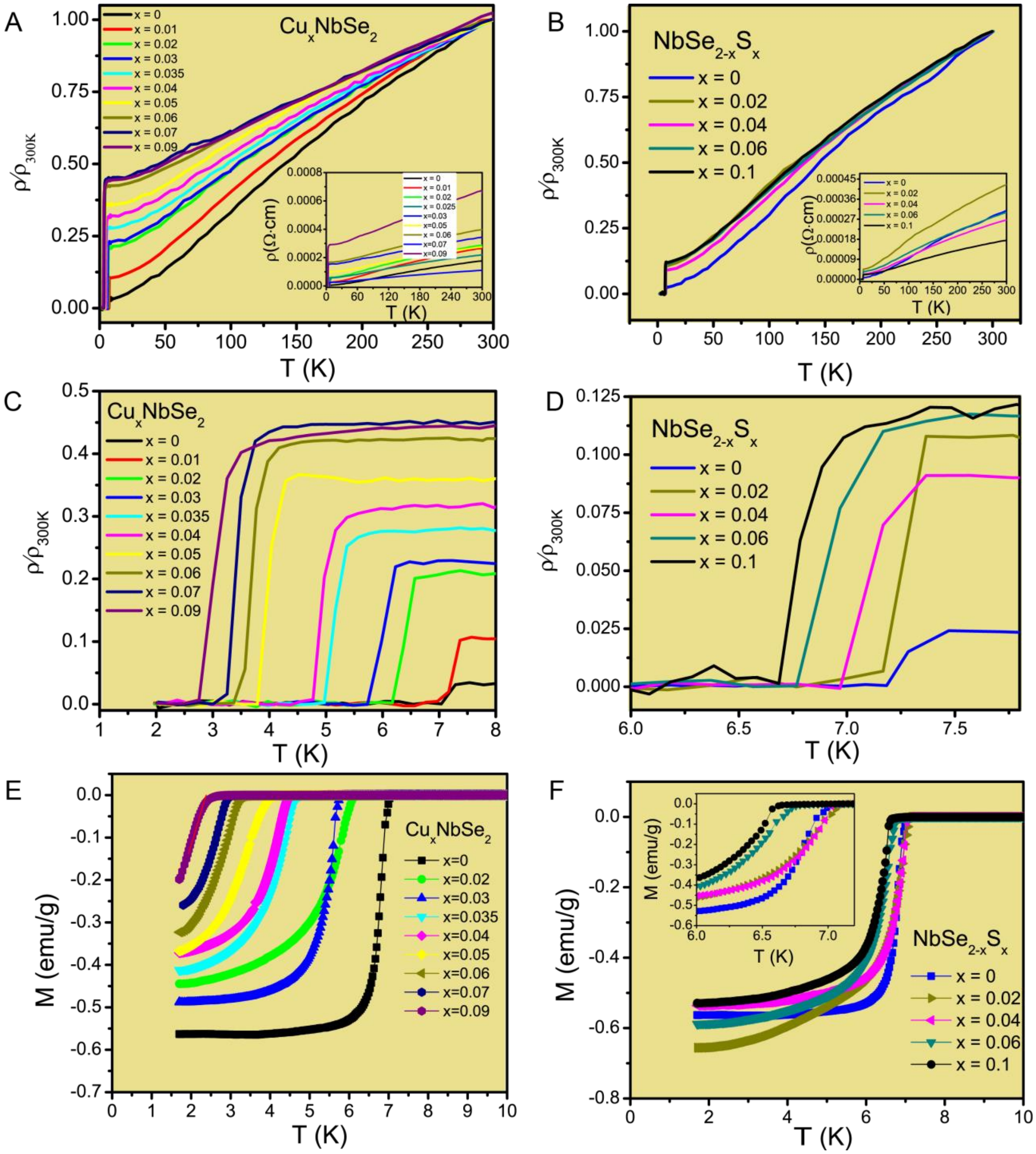

**Fig. 3.**

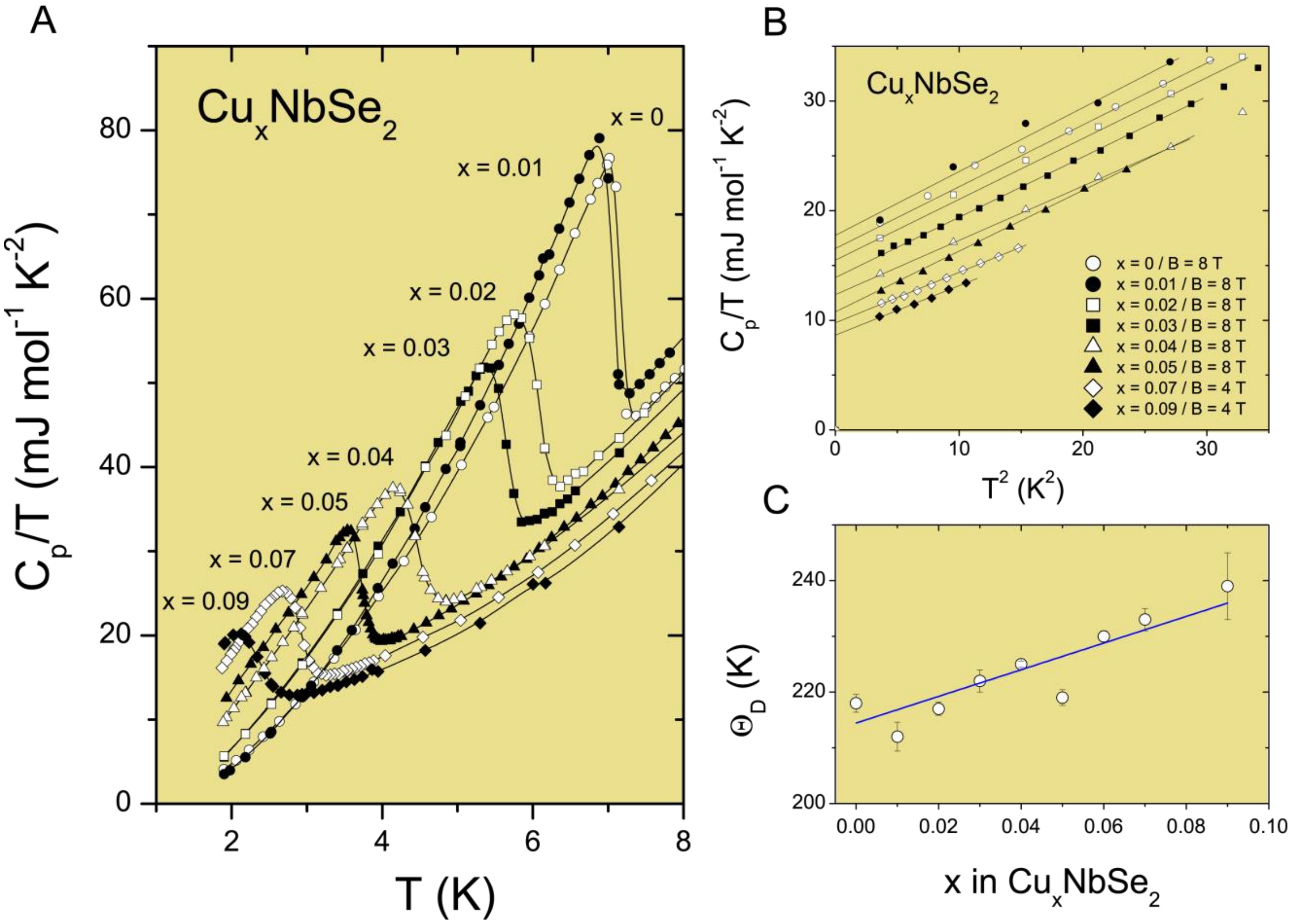

A
B
C
$Cu_xNbSe_2$
x = 0
x = 0.01
x = 0.02
x = 0.03
x = 0.04
x = 0.05
x = 0.07
x = 0.09
$C_p/T$ (mJ mol$^{-1}$ K$^{-2}$)
T (K)
x = 0 / B = 8 T
x = 0.01 / B = 8 T
x = 0.02 / B = 8 T
x = 0.03 / B = 8 T
x = 0.04 / B = 8 T
x = 0.05 / B = 8 T
x = 0.07 / B = 4 T
x = 0.09 / B = 4 T
$T^2$ (K$^2$)
$\Theta_D$ (K)
x in $Cu_xNbSe_2$

**Fig. 4.**

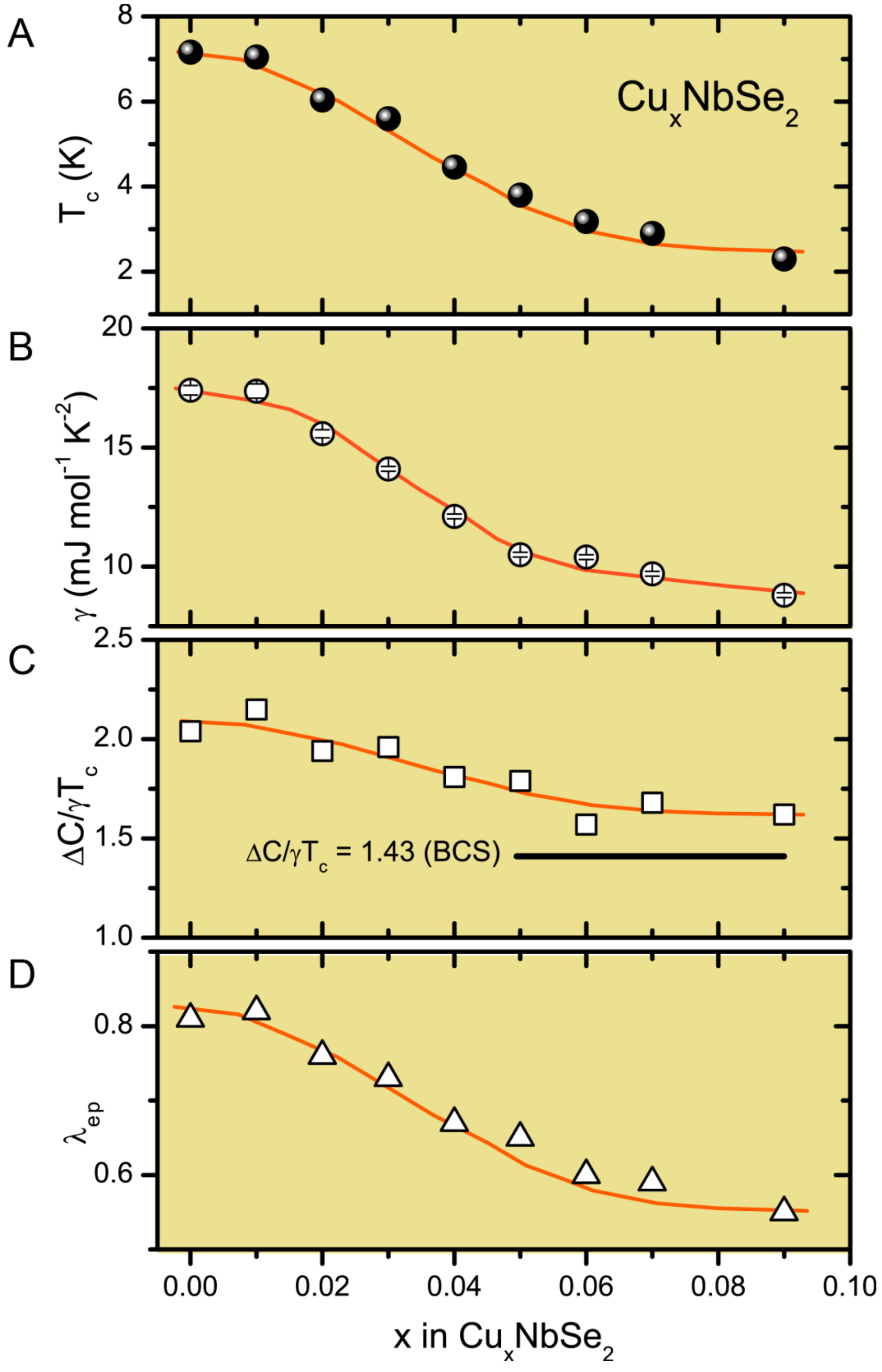

**Fig. 5.**

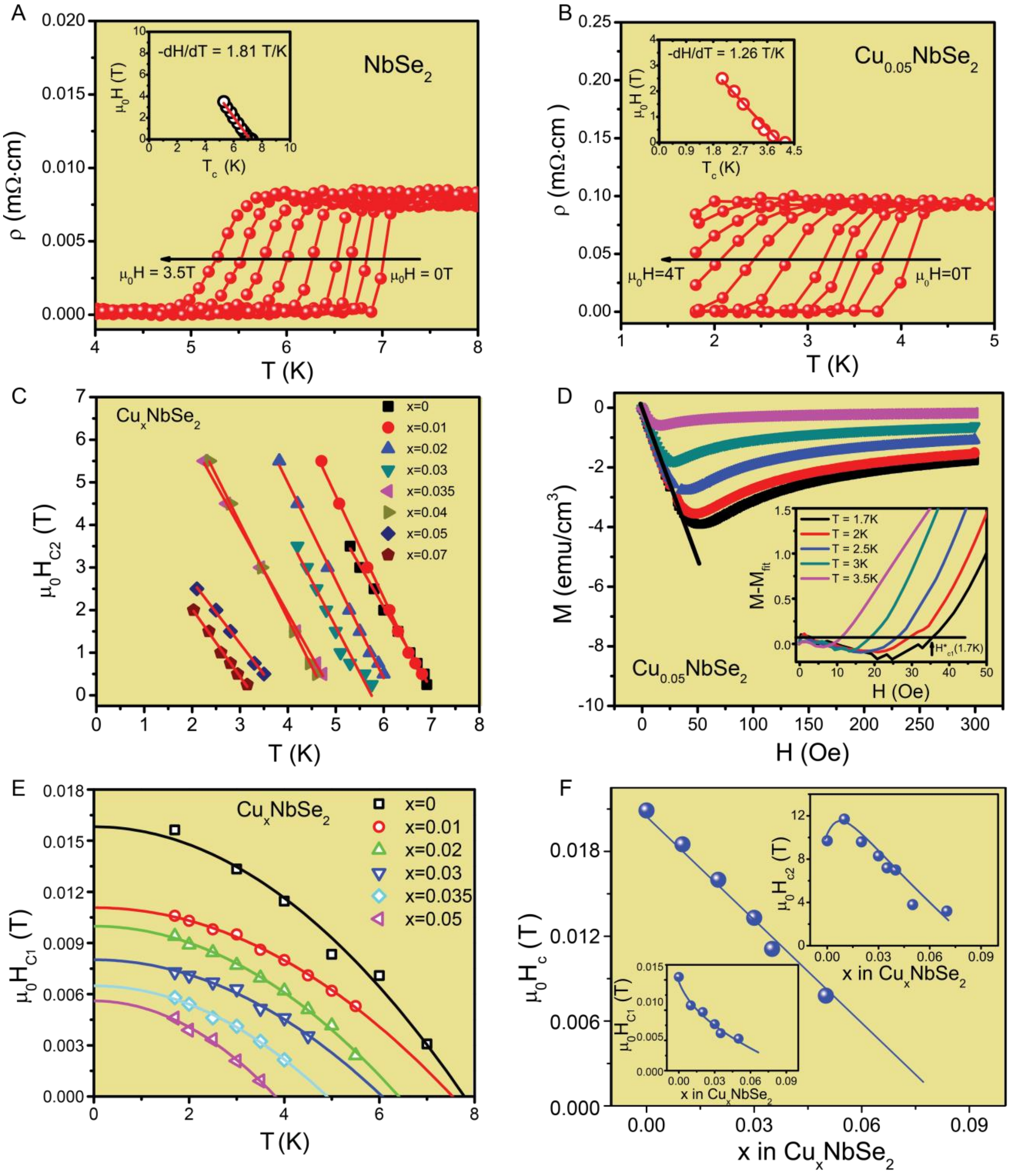

**Fig. 6.**

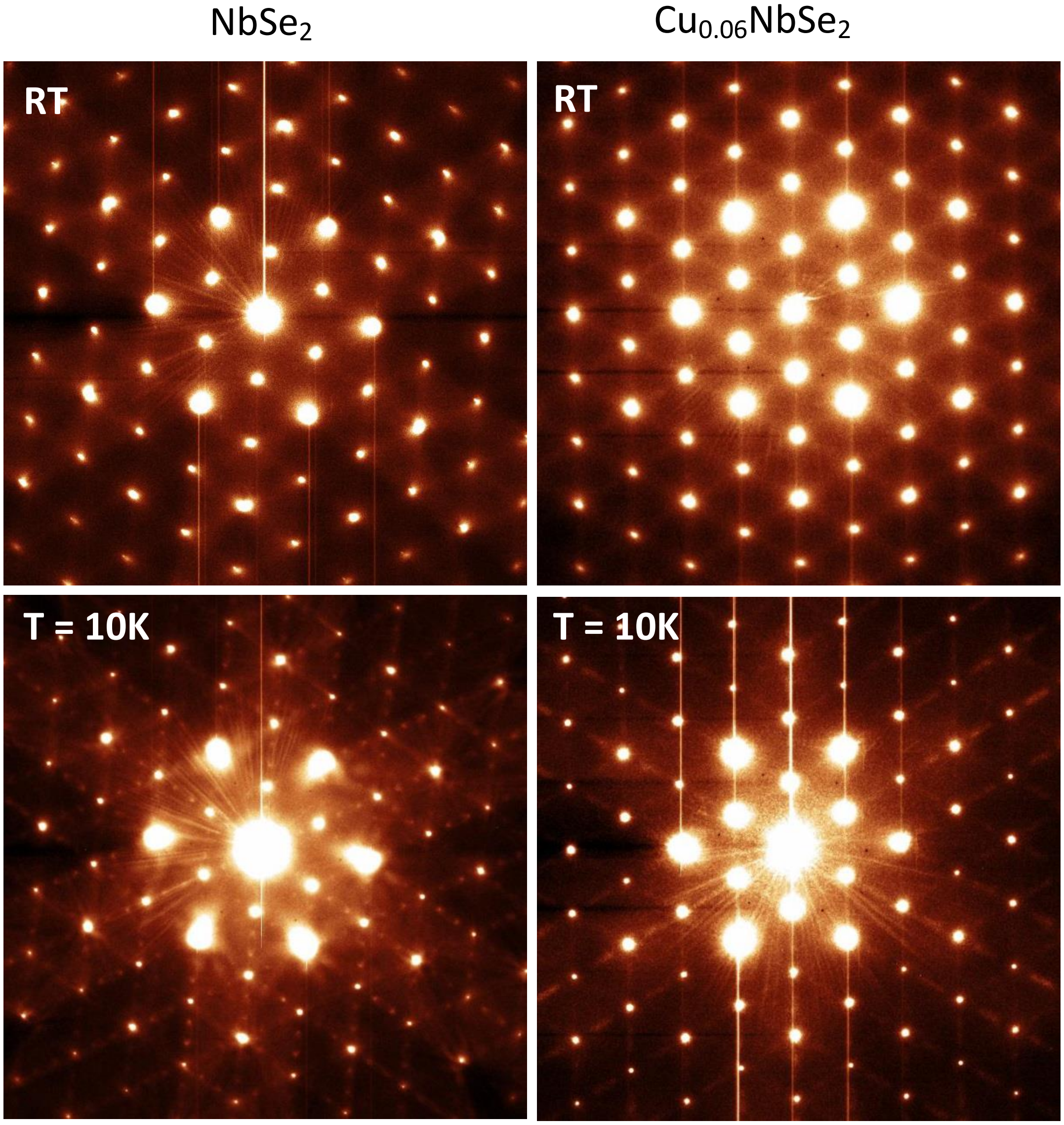

**Fig. 7.**

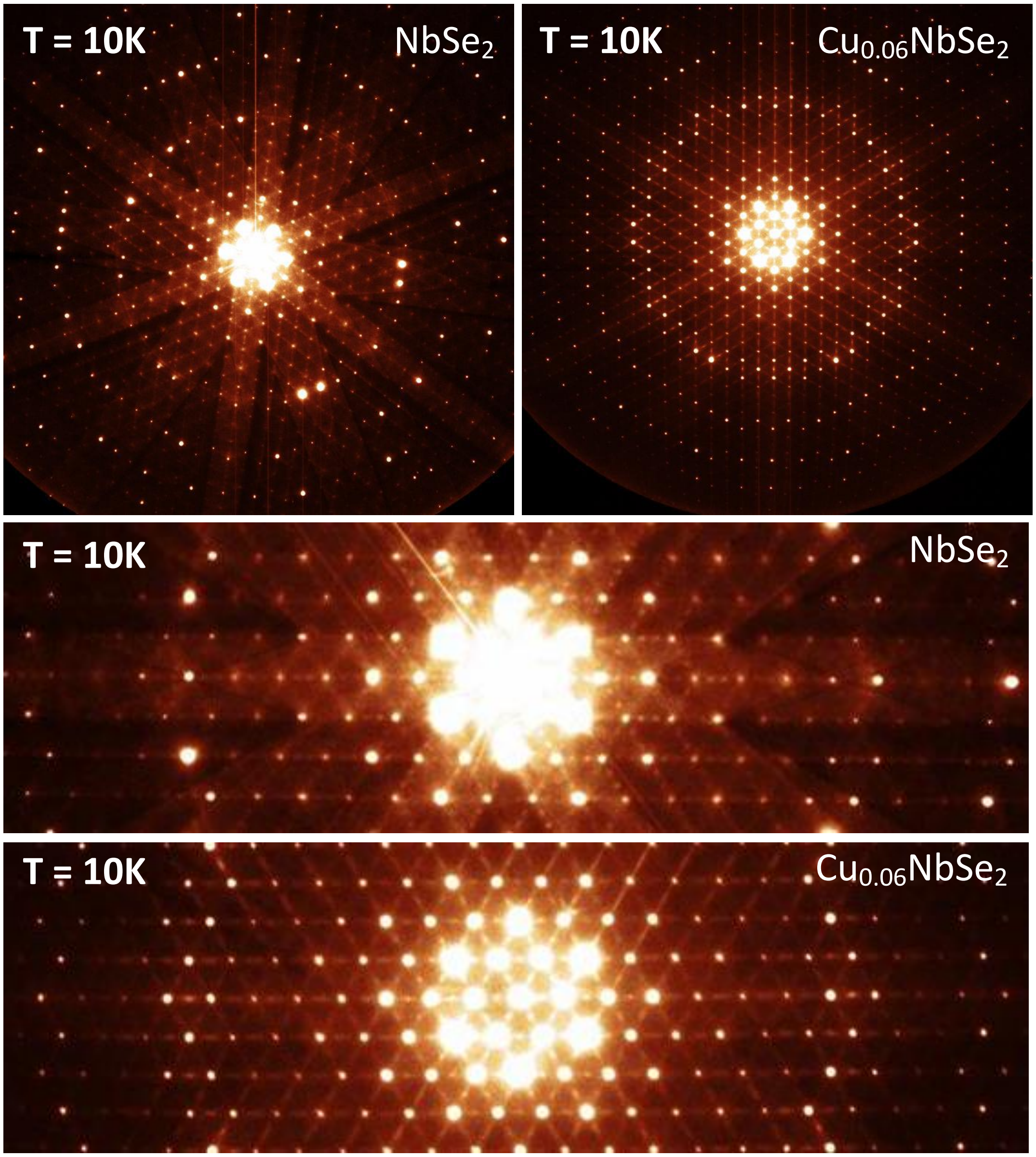

Fig. 8

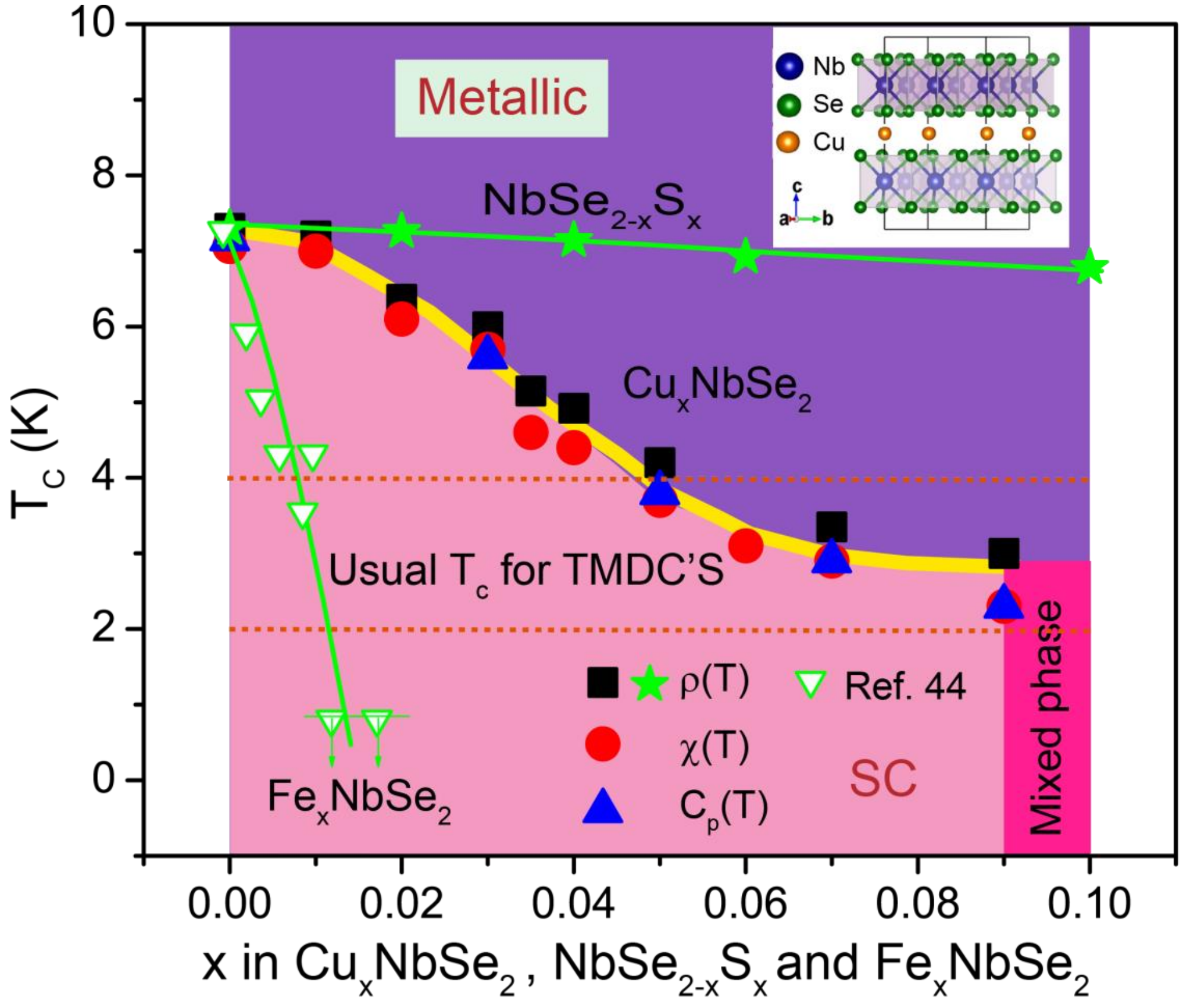